\documentclass{jfm}
\usepackage{graphicx}
\usepackage{epstopdf, epsfig}
\usepackage{newtxtext}
\usepackage{newtxmath}
\usepackage[x11names, svgnames, dvipsnames]{xcolor}
\usepackage{soul} 
\usepackage{float}
\usepackage{lineno}
\usepackage[normalem]{ulem}

\graphicspath{{./figures/}}

\newcommand{\mean}[1]{\langle #1 \rangle}

\newcommand\Fro{\mbox{\textit{Fr}}}

\shorttitle{Internal gravity waves in flow past a bluff body under different levels of stratification}
\shortauthor{Gola et al.}

\title{Internal gravity waves in flow past a bluff body under different levels of stratification}

\author{Divyanshu Gola\aff{1}, S. Nidhan\aff{1}, S. Sarkar\aff{1}\corresp{\email{ssarkar@ucsd.edu}}}

\affiliation{\aff{1}Department of Mechanical and Aerospace Engineering, University of California San Diego, CA 92093, USA}	

\begin{document}
\maketitle
\nolinenumbers
\begin{abstract}

The flow field of a bluff body, a circular disk,  that moves horizontally in a stratified environment is studied using large eddy simulations (LES).  Five levels of stratification (body Froude numbers of $\Fro = 0.5, 1, 1.5, 2$ and $5$) are simulated at Reynolds number of $\Rey = 5000$ and {Prandtl number of $Pr =1$}. {A higher $\Rey = 50,000$ database at $\Fro = 2, 10$ and $Pr =1$ is also examined for comparison.}  The wavelength and amplitude of steady lee waves are compared with a linear-theory analysis. Excellent agreement  is found over the entire range of $\Fro$  if an `equivalent body' that includes the  separation region is  employed for the linear theory. For asymptotically large distance, the velocity amplitude varies theoretically as $\Fro^{-1}$  but a correction owing to  dependence of the separation zone on $\Fro$ is needed. The wake waves propagate in a narrow band of angles with the vertical and have a wavelength that increases with increasing $\Fro$. The envelope of wake waves, demarcated using buoyancy variance, exhibits self-similar behavior. The higher $\Rey$ results are consistent with the  buoyancy effects exhibited at the lower $\Rey$. The wake wave energy is larger at  $\Rey = 50,000$. Nevertheless, independent of $\Fro$ and $\Rey$, the ratio of the wake wave potential energy to the wake turbulent energy  increases to approximately $0.6 \sim 0.7$  in  the nonequilibrium (NEQ) stage  showing their energetic importance besides suggesting universality in this statistic. There is a crossover of energetic dominance of lee waves at $\Fro <2$ to wake-wave dominance at $\Fro \approx 5$.

\end{abstract}

\section{Introduction}

\subsection{Internal waves and their impact}

Relative motion between a submerged body and its stratified environment gives rise to different types of internal gravity waves. Observations of these  waves in  the ocean and the atmosphere prompted the first studies of these waves in the geophysical context. Flow over mountains gives rise to lee waves that can lead to changes in properties of air, sometimes visualised as wave clouds. In the ocean,  substantial energy is  transferred from impinging mean currents and tides at bottom topography to internal waves e.g \cite{GarrettK2007} which, through nonlinear processes break down to turbulence, e.g. \cite{SarkarS2017}, and the associated turbulent mixing  provides an important control on ocean stratification and currents, e.g. \cite {WunschF2004}. The motivating application - bodies moving in a stratified environment - of the present work is different. Here, the body generates lee waves, which are steady in the reference frame of the body,  and the unsteady motions in the wake also generate waves (unsteady in the frame of the body). Characterization of the space-time structure and energetics of these waves is of interest  as is quantification of their relationship with the wake and the body generator.

The internal or body Froude number, $\Fro = U / ND$, is the important overall parameter that governs internal gravity waves and buoyancy effects on wake dynamics. Here, the density-stratified background is characterized by the buoyancy frequency $N$ where $N^2 = - (g/\rho_0) \partial \rho_b /\partial z$,   the  relative velocity between the  body and the ambient by $U$ and the size of the body by $D$. In geophysical flows, $\Fro$ is generally less or much less than $O(1)$, while in engineering applications (underwater submersibles, aerial vehicles, wind farms, marine turbines) $\Fro \geq O(1)$.

\subsection{{Lee wave studies in the laboratory  and  with simulations}}

 Laboratory experiments  of stratified flow past a model hill \citep{hunt_experiments_1980} and model ridges \citep{castro_stratified_1983} have characterized lee waves  through extensive flow visualization. \cite{chomaz_structure_1993} experimentally studied and classified the change of near-wake structure of a sphere  wake in uniform stratification as $\Fro$ is decreased from greater to less than $O(1)$ values. In a companion study, \cite{bonneton_internal_1993} described  wave characteristics on horizontal planes above the wake  using isopyncal displacement inferred from fluorescent dye images.  Nonuniform stratification was treated by \cite{robey_thermocline_1997} who, using experiments and numerics,  found  the wavefield of a sphere moving at the base of a  thermocline  to be composed of discrete modes that give rise to horizontal-plane patterns reminiscent of Kelvin wakes on the air-sea boundary.  Recently,  \cite{meunier_internal_2018} investigated the wavefield of bodies of different shapes and determined how the wavelength and velocity amplitude of lee waves varied with $\Fro$. They also proposed a model for lee wave amplitude that involves modeling the drag  as a point force in the Navier-Stokes equations.
 
  Various numerical simulations have also described these lee waves and their effects on the flow at the body and in the wake.  For a sphere at $\Rey =200$, \cite{hanazaki_numerical_1988} found that lee waves suppress separation on decreasing $\Fro$ when $\Fro > 1$ but induce separation on decreasing $\Fro$ when $\Fro < 1$. Lee waves  lead to oscillatory modulation of the  wake width and velocity \citep{pal_direct_2017}  and also, near the body, lead to a local minimum of the wake defect velocity at $Nt = \pi$ corresponding to the half-wavelength of the lee wave \citep{chongsiripinyo_decay_2020}. {Here, $Nt = Nx/U = x/(\Fro D)$ converts distance behind the body to time in buoyancy units.} Lee waves have a stronger effect on the wake of a slender body, i.e.,  large value of   aspect ratio AR given by $L/D$ where $L$ is the length of the body. When  the half-wavelength of the lee wave matches $L$, a condition that is realized at  a critical value of $\Fro_{c} = AR/\pi$, the wave leads to maximal contraction of the separated flow at the trailing edge. Thus,  in an LES study \citep{ortiz_spheroid_2019} of flow past  a 4:1 spheroid at $\Rey = 10^4$, the wake which was turbulent at high $\Fro$ exhibited strong suppression of wake turbulence at $\Fro = 1$. The problem of a disk at $\Rey = 5000$  in nonlinear stratification with minimum $\Fro =1$ was studied by \cite{gola_pycnocline_2023}.  Linear theory \citep{voisin_internal_1_1991, voisin_iwb_2003, voisin_sphere_2007} was used with an equivalent body comprising of the disk and its separation bubble  and found to predict the wave amplitude of a single test case ($\Fro =1$) with linear stratification but not the nonlinearly stratified cases. In the present work,  lee waves in linear stratification will be characterized over a wide range of    $\Fro$  and the applicability of linear theory will be evaluated.

\subsection{{Evolution of stratified turbulent wakes}}

Turbulent stratified wakes exhibit a stark contrast in the evolution of their deficit velocity  and length scales relative  to their unstratified counterpart.  Although the present focus is on internal waves,  previous work in this area is briefly summarized for completeness. 
  
Several experimental studies \citep{lin_wakes_1979, SpeddingBF_jfm:1996,Lin:1992turbulent,spedding_evolution_1997, bonnier_experimental_2002} have reported significant buoyancy effects on the  evolution of wake thickness and/or wake deficit in the flow behind a horizontally moving sphere. \cite{spedding_evolution_1997} found that the decay of the wake deficit velocity is characterised by three stages: {the near wake regime, the non-equilibrium (NEQ) regime, where the wake decay rate is reduced as it adjusts to the surrounding stratification, and the {quasi-two dimensional  (Q2D)} regime, where pancake vortices form and the wake decay speeds up. The transition between stages  occurred at nominally fixed values of $Nt$. In another laboratory study, \cite{spedding_anisotropy_2001} showed that stratification biases turbulence towards a faster decay of the vertical fluctuations as compared to the horizontal fluctuations. }

{Numerical investigations  of the problem started with a temporal model \citep{gourlay_numerical_2001} where the wake is simulated in a frame moving with the body,  a streamwise periodic domain is used and flow statistics evolve in time. The model does not include the wake generator and is  instead initialised with a flow field  that approximates experimentally observed mean velocity and turbulence intensities at some distance from the body. Since the boundary layer on the body and the small scales in the near wake do not need to be resolved,  the temporal model has the advantage that the simulated flow can progress into the late wake with reasonable computational cost.  The limitations are that the absence of the wake generator excludes lee waves and their influence on the wake as well as vortex shedding from the body. Also, temporal models are  sensitive to initial flow conditions \citep{redford_universality_2012}. The alternative approach of body-inclusive simulations will be adopted here. }

The multistage decay of the wake has been studied numerically using temporal models \citep{gourlay_numerical_2001,dommermuth_numerical_2002,brucker_comparative_2010,Diamessis2011, de_stadler_simulation_2012,zhou_large_2019}.The anisotropy created by stratification suppressing vertical fluctuations leads to decreased turbulent production in the wake \citep{brucker_comparative_2010}, leading to the longer lifespan of the stratified wake, a result that is supported by \cite{redford_numerical_2015} in their DNS of a weakly stratified turbulent wake. The buoyancy-induced decrease in the correlation coefficient between vertical and horizontal fluctuations is a generic feature of stratified shear flows and was identified  by \cite{JacobitzSV:1997} in homogenous shear flow and also by others in following studies of a variety of shear flows.

In recent times, body-inclusive simulations, e.g.  \cite{orr_numerical_2015, pal_direct_2017, ortiz_spheroid_2019, chongsiripinyo_decay_2020} have been conducted to increase the realism of simulations. Such an approach captures flow separation, vortex shedding from the body and also the lee waves leading to more accurate representation of the near and intermediate wake. {Body inclusive simulations are akin to lab experiments by virtue of including the body, while also alleviating the aforementioned limitations of temporal-model simulations.} Besides, body-inclusive simulations leads to statistically steady data so that numerical techniques like spectral proper orthogonal decomposition (SPOD) can be fruitfully used to extract wake structures that are statistically space-time coherent, for example \cite{nidhan_spectral_2020,nidhan_analysis_2022}. Reaching beyond $x/D = O(100)$ in body-inclusive simulations is computationally too expensive but this issue can be overcome by a hybrid simulation technique that continues the flow to larger $x/D$ (equivalently $Nt$) in a followup simulation that uses a temporal model  \citep{Pasquetti:2011}  or a spatial model \citep{VanDine2018} with a coarser grid that needs to resolve only the large length scales of the far wake.

\subsection{{Wake waves and their link to wake structures}}

Wake waves are internal waves forced by the unsteady fluctuations in the turbulent wake.
The first report of   stratified wake  waves is by \cite{gilreath_experiments_1985} who observed "short, random" unsteady waves in their experiments with a streamlined body in towed and self-propelled modes. Besides uniform stratification, they also considered a pycnocline to find solitary internal wave propagation at low $\Fro$. \cite{bonneton_internal_1993}  distinguished wake waves from lee waves and,  by spectral analysis of density-probe measurements, linked these waves to coherent wake structures.  \cite{robey_thermocline_1997}  linked the  wake waves to the size of large-scale structures in the wake. 
Towed sphere experiments  by \cite{brandt_internal_2015} showed that lee waves dominate in the $\Fro \lesssim 1$ regime while wake waves dominate in the $\Fro \gtrsim 1$ regime. They also provide a qualitative estimate of the total potential energy that was found to scale as $\Fro^{2}$. The recent work of \cite{meunier_internal_2018} considered wake waves in addition to lee waves and quantified the dependence of their wavelength and velocity amplitude on $\Fro$. 

\cite{abdilghanie_internal_2013} conducted a comprehensive study of  the internal wave field of a stratified turbulent wake using temporal simulations for a wide range of $\Rey$ and found that, at higher $\Rey$, there is a broader range of wave propagation angles ($40^{\circ} - 55^{\circ}$), at least at early $Nt$, relative to lower values of $\Rey$. \cite{brucker_comparative_2010} studied  towed and self-propelled wakes using temporal simulations at $\Rey = 5 \times 10^{4}$ and various $\Fro$. They concluded that internal wave flux dominated the turbulent dissipation in the wake kinetic energy budget for $20 < Nt < 75$. This was also supported by \cite{rowe_internal_2020} who, in their temporal-model simulations over a broad range of $\Rey$, $\Fro$, found that internal wave radiation is an important sink for wake kinetic energy after $Nt = 10$. Indeed, the energetic importance of  internal waves for turbulence in  stratified shear flows is a more general result in view of such demonstrations for a shear layer \citep{pham_dynamics_2009, Watanabe2018} and a boundary layer \citep{taylor_internal_2007}.

\subsection{{Objectives of present study}} \label{objectives}

{Body inclusive simulations capture  lee waves as well as wake waves and present an opportunity to characterize and contrast their properties. This motivates the present waves-focussed study of a disk wake in linear stratification at $\Rey = 5000$. 

 Our results on scaling laws for $\Fro$ dependency of internal waves in turbulent  flow past a disk will be compared  with  previous laboratory experiments and, for wake waves,  also with body exclusive temporal model simulations. The skill of linear theory in predicting lee waves will be assessed. The internal wave field in cases from a recent study of disk wakes  at an order of magnitude larger $\Rey = 50,000$ that focussed on wake decay laws  will also be analyzed.

The specific questions that we intend to answer as stratification is varied parametrically for a turbulent wake are as follows: (1) How does linear theory applied to an equivalent  body, an approach which was shown to  predict body-generated lee waves in a turbulent disk wake at a single $\Fro =1$, work over the range of $\Fro$ considered here? For example, how do  linear theory results,  $ w \propto \Fro^{-1}$ scaling of the amplitude and $\Lambda = 2\pi \Fro D$ scaling of wavelength, compare with the simulation results? (2)  How do internal wave structural properties (phase angle, wavelength, spatial distribution) change as a function of $\Fro$? Are there scaling laws for the $\Fro$ dependence? (3) There is fluctuation energy in the turbulent wake and also outside it in the form of internal waves. How are kinetic and potential energy partitioned between the wake and wave field?  How does the energy in wake waves compare with that in lee waves? (4) Are results for the two cases at higher $\Rey = 50,000$ consistent with trends exhibited in the parametric study at  the lower value of $\Rey = 5000$?
}

A disk is used as the wake generator in body inclusive large eddy simulations, the numerics of which are laid out in \S \ref{methodology}. Lee waves and their dependence on $\Fro$ is described in \S \ref{lee_waves}. Characteristics of wake generated internal gravity waves are described in \S \ref{unsteady_waves}. Energy partition of the entire flow field is done to calculate the energies associated with lee waves, wake waves, mean wake and turbulent wake, and their variation with $\Fro$ is analysed in \S \ref{energetics}. {The influence of Reynolds number is briefly assessed  in \S \ref{comparison}}. To conclude, a summary and a discussion are presented in \S \ref{conclusions}.

\section{Methodology}\label{methodology}

\subsection{Governing equations and numerical scheme}\label{numerical_details}

\begin{figure}

\centerline{\includegraphics[width=1.0\linewidth, keepaspectratio]{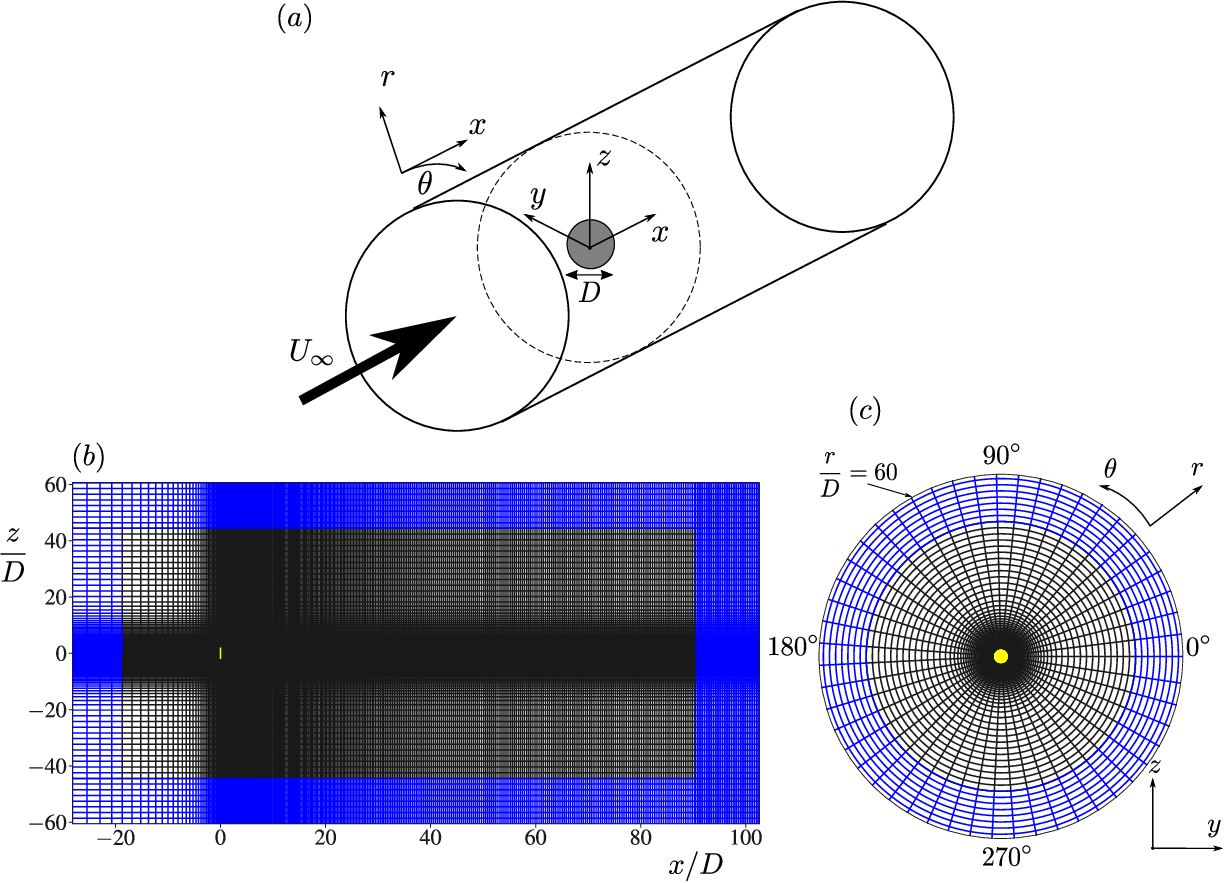}}
\caption{{(a) Schematic showing the simulation setup and domain. (b) Radial and streamwise grids visualised on $y = 0$ plane. (c) Radial and azimuthal grids visualised on $x = 0$ plane. (Every fifth gridline is shown, with blue lines representing sponge region. Disk (not to scale) is shown in yellow.)}}
\label{fig:simsetup}
\end{figure}

The wake of a disk is simulated for five different cases of stratification by solving the three dimensional, incompressible, unsteady form of the conservation equations for mass, momentum and density. A high-resolution large eddy simulation (LES)  with the Boussinesq approximation for density effects is used. The disk,  with diameter $D$ and {thickness $0.01D$},  is immersed perpendicular to a flow with velocity $U_{\infty}$. The equations are numerically solved in cylindrical coordinates but both Cartesian $(x,y,z)$ and cylindrical $(r,\theta,x)$ coordinates are  used as appropriate in the discussion. Here, $x$ is streamwise, $y$ is spanwise and positive along $\theta=0^{\circ}$, and $z$ is vertical and positive along $\theta=90^{\circ}$ (Figure \ref{fig:simsetup}). The density field $(\rho(\mathbf{x},t))$ is split into a constant reference density $(\rho_{0})$, the variation of the background $(\Delta \rho_{b}(z))$, and the flow induced deviation, $(\rho_{d}(\mathbf{x},t))$ so that $\rho(\mathbf{x},t)=\rho_{0}+\Delta \rho_{b}(z)+\rho_{d}(\mathbf{x},t)$. The Reynolds number of the flow, defined as $\Rey = U_{\infty} D/\nu $ is $5000$. The different levels of stratification are quantified by the Froude number, $Fr = U_{\infty}/ ND$, which takes values of $0.5, 1, 1.5, 2$ and $5$ for the five cases. Here, $N$ is the buoyancy frequency given by $N = \sqrt{(-g/\rho_{0}) (\partial \Delta \rho_{b} (z) / \partial z) }$.

The filtered nondimensional equations, {using $D$, $U_{\infty}$ and $\rho_{0}$ as the characteristic length, velocity and density scales respectively,} are as follows 

\begin{equation} 
\frac{\partial u_{i}}{\partial x_{i}} = 0,
\label{conservation_eqn}
\end{equation}

\begin{equation} 
\frac{\partial u_{i}}{\partial t} + u_{j}\frac{\partial u_{i}}{\partial x_{j}} = -\frac{\partial p}{\partial x_{i}} + \frac{1}{\Rey} \frac{\partial}{\partial x_{j}}\Big[(1 + \frac{\nu_{sgs}}{\nu}) \frac{\partial u_{i}}{\partial x_{j}}\Big] - \frac{\rho_{d}}{Fr^{2}} \delta_{i3}, 
\label{momentum_eqn}
\end{equation}

\begin{equation} 
\frac{\partial \rho}{\partial t} + u_{j}\frac{\partial \rho}{\partial x_{j}} = \frac{1}{\Rey Pr} \frac{\partial}{\partial x_{j}}\Big[(1 + \frac{\kappa_{sgs}}{\kappa})\frac{\partial \rho}{\partial x_{j}}\Big], 
\label{density_eqn}
\end{equation}
{where $u_{i}$ refers to the filtered nondimensional velocities in the cartesian coordinate system for $i = 1, 2$, and $3$ respectively.}  {The Smagorinsky eddy viscosity model with dynamic procedure~\citep{germano_dynamic_1991} is employed to obtain $\nu_{sgs}$ wherein, the coefficient $C$ in the expression $\nu_{sgs} = C \tilde{\Delta}^{2}|\tilde{S}|$ ($\tilde{\Delta}^{3}$ being the cell volume and $\tilde{|S|}$ being the strain rate magnitude) is obtained dynamically using the least squares approach of \cite{lilly_proposed_1992} and the Lagrangian averaging method of \cite{meneveau_lagrangian_1996}. The Lagrangian averaging method takes the cumulative average of $C$ as a function of time, with more weight assigned to recent times in flow evolution.} The Prandtl number ($Pr=\nu / \kappa$) as well as the subgrid Prandtl number ($Pr=\nu_{sgs} / \kappa_{sgs}$) is set to unity. {Second order central finite difference is used in space.  A low storage third-order Runge-Kutta scheme for the advective terms and a Crank-Nicolson term for the viscous terms are used to advance the solution in time. } The disk is represented using the immersed boundary method of \cite{balaras_modeling_2004} and \cite{yang_embedded-boundary_2006}.  

\subsection{Domain and Statistics}

{The domain extends from $x/D = -L_{x}^{-} = -30$ to $x/D =  L_{x}^{+} =102$ in the streamwise direction and from $r/D = 0$ to $r/D =  L_{r} =60$ in the radial direction (figure \ref{fig:simsetup} b,c). The number of grid points in the streamwise, azimuthal and radial direction are $N_{x} = 2176$, $N_{\theta} = 256$ and $N_{r} = 479$. }
{The disk surface  is resolved into 47312 triangles to identify the fluid-solid interface.}{ The streamwise and radial grids in the flow solver are structured and employ uniform stretching as distance from the disk increases. There are 73 points along the disk in the radial direction. {Maximum values of $\Delta x/\eta$ and $\Delta r / \eta$, where $\eta = (\nu^{3} / \varepsilon)^{1/4}$ is the Kolmogorov length scale, are used to ascertain adequate grid resolution.  The chosen grid has $(\Delta x / \eta)_{max} = 4.84$ and $(\Delta r / \eta)_{max} = 5.43$ across all cases.} The grid in the azimuthal direction is homogenous. } {Owing to the high grid resolution, the subgrid viscosity ($\nu_{sgs}$) is not large. The instantaneous maximum value of subgrid viscosity fluctuates between $\nu_{sgs} / \nu = 2.5$ and $\nu_{sgs} / \nu = 5$ after statistical stationarity is achieved. The location of this maximum value is $0.5 < r/D < 1.5$ and $0< x/D < 2$. Beyond $x/D = 10$, $\nu_{sgs}/\nu < 1$.} Statistics are collected by temporal averaging after the initial transient has subsided and statistical steady state has been reached. The time interval for averaging is $140 D / U_{\infty}$. For any dependent variable in the simulation, $\langle . \rangle$  represents the time average and superscript $'$  represents the fluctuation about that average. {From this section onwards, all the variables will appear in their dimensional form and any non-dimensionalisation will be explicitly shown, e.g. $x/D$ or $w/U_{\infty}$.}

\subsection{Boundary Conditions}
For boundary conditions, the inlet ($x/D = -30$) has a uniform freestream of magnitude $U_{\infty}$ in the positive $x$ direction, the outlet ($x/D = 102$) has an Orlanski-type convective boundary condition \citep{orlanski_simple_1976}, and the radial boundary ($r/D = 60$) has a Neumann boundary condition for density as well as velocity. {The centerline boundary ($r = 0$) is handled by taking the average of the two symmetrically located ghost cells over the centerline to define the radial and azimuthal velocity components at the centerline.} {To prevent spurious reflection of waves back into the domain, sponge layers are used at the three boundaries of the computational domain. The radial sponge starts at $r/D = 45$ and goes upto the radial boundary $r/D = 60$, the inlet sponge is from $x/D = -30$ to $x/D = -20$ and the outlet sponge is from $x/D = 90$ to $x/D = 102$. Each of the sponge layers use a quadratic damping function of the form $f_{damp} = C ( (x - x_{spng})/(L_{x} - x_{spng}) )^{2} (\phi_{0} - \phi(x))$, where the variable $\phi(x)$ is relaxed to the target value $\phi_{0}$  in the sponge layer given by $x_{spng} < x < L_{b}$. For example, in the radial sponge, where $x$ in $f_{damp} (x)$ is equivalent to $r$, the sponge begins at $r_{spng}/D = 45$ and extends up to $L_{r}/D = 60$, and the strength of the damping function is $C = 5$. }

\section{Body generated lee waves}\label{lee_waves}

\begin{figure}

\centerline{\includegraphics[width=\linewidth, keepaspectratio]{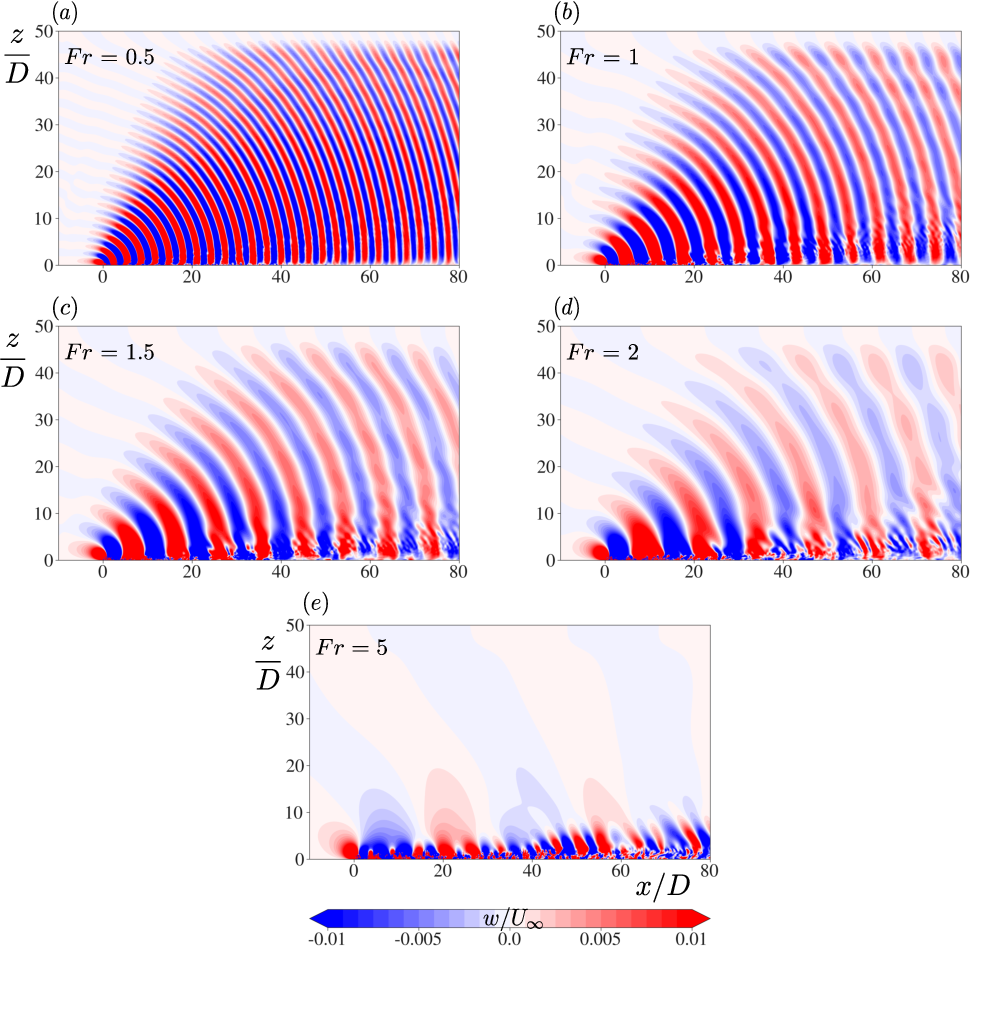}}
\caption{ Instantaneous contours of vertical velocity showing body-generated lee waves (steady in the simulation frame)  on the $\theta = 90^{\circ}$, vertical center-plane for the five cases. 
}
\label{fig:lee_waves}
\end{figure}

In stratified environments, the wave component of the flow field consists of body generated lee waves and wake generated internal waves. The former is steady in time with respect to the body while the latter is unsteady. Figure \ref{fig:lee_waves} illustrates these two types of waves for the five values of  Froude number by plotting the vertical velocity at the top half of the vertical center-plane. The steady lee waves are seen as the large band of waves spanning most of the domain while the unsteady waves are seen as a radially thinner band of irregular waves superposed on the lee waves. \cite{gola_pycnocline_2023} studied the steady lee waves generated by a disk at $\Rey = 5000$ focussing on how nonlinear stratification (pycnocline) as well as the relative position of the pycnocline with respect to the disk modifies the flow. The minimum background Froude number for the pycnocline cases  was $Fr =1$ as was the constant value of $Fr$ in the linearly-stratified baseline case.

\begin{figure}

\centerline{\includegraphics[width=0.75\linewidth, keepaspectratio]{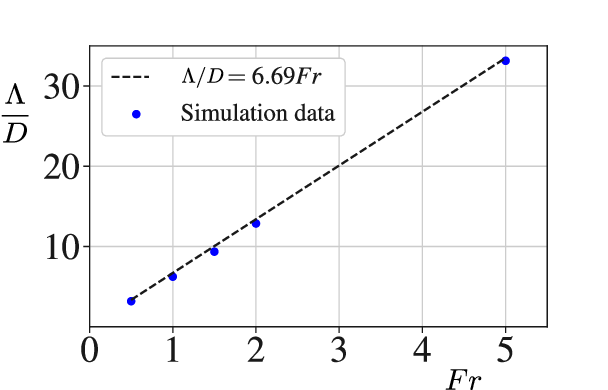}}
\caption{ Wavelength of body-generated lee waves.}
\label{fig:lw_wl}
\end{figure}

\begin{figure}

\centerline{\includegraphics[width=\linewidth, keepaspectratio]{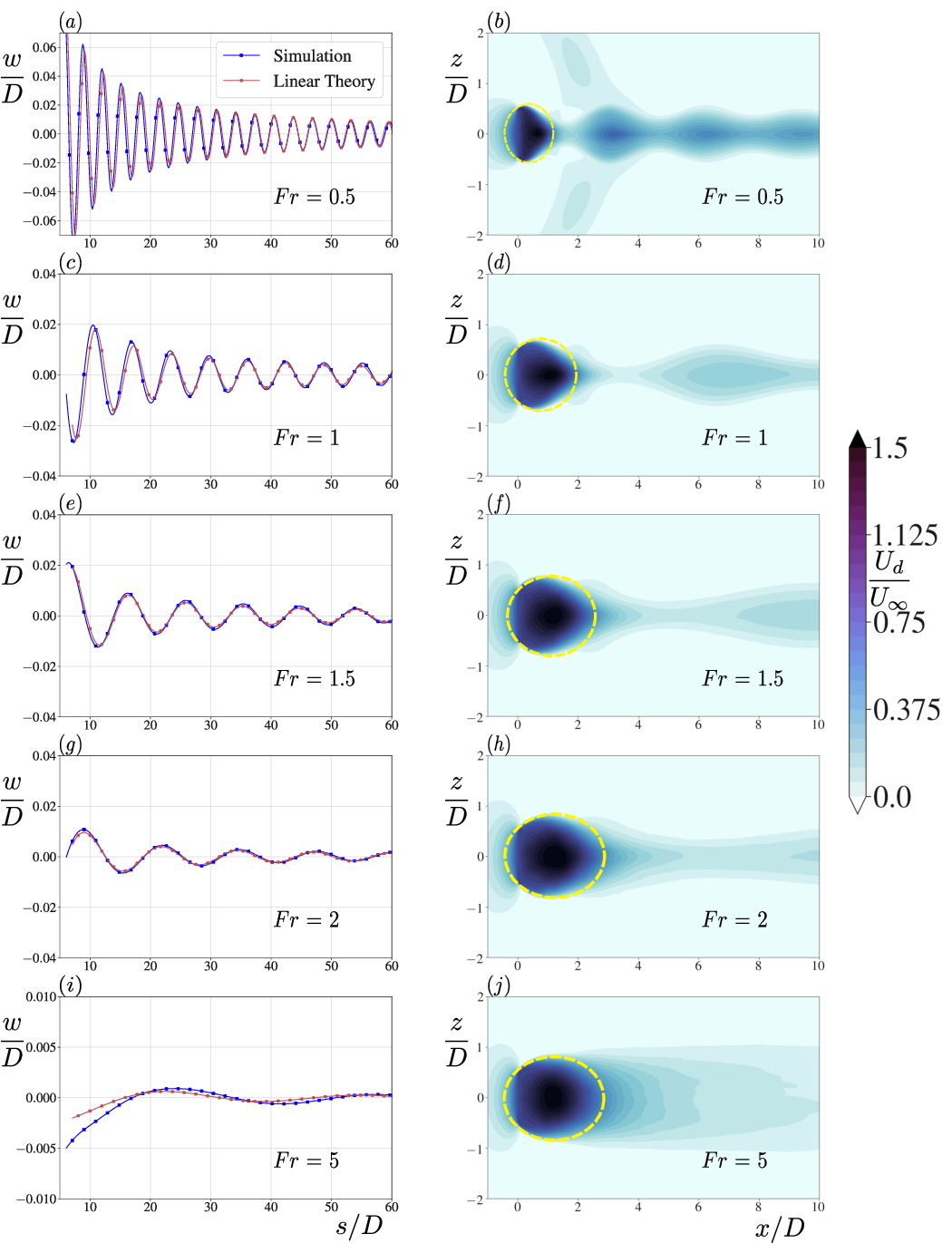}}
\caption{ Body-generated lee waves (a,c,e,g,i): vertical mean velocity plotted along the line $z=x, y=0$ on the top half vertical plane ($s = \sqrt{x^{2} + z^{2}}$). {(b,d,f,h,j): defect velocity ($U_{d}$) contours showing the increasing size of the fitted Rankine ovoid with increasing \Fro.}}
\label{fig:separation_bubble}
\end{figure}

\begin{figure}

\centerline{\includegraphics[width=1.1\linewidth, keepaspectratio]{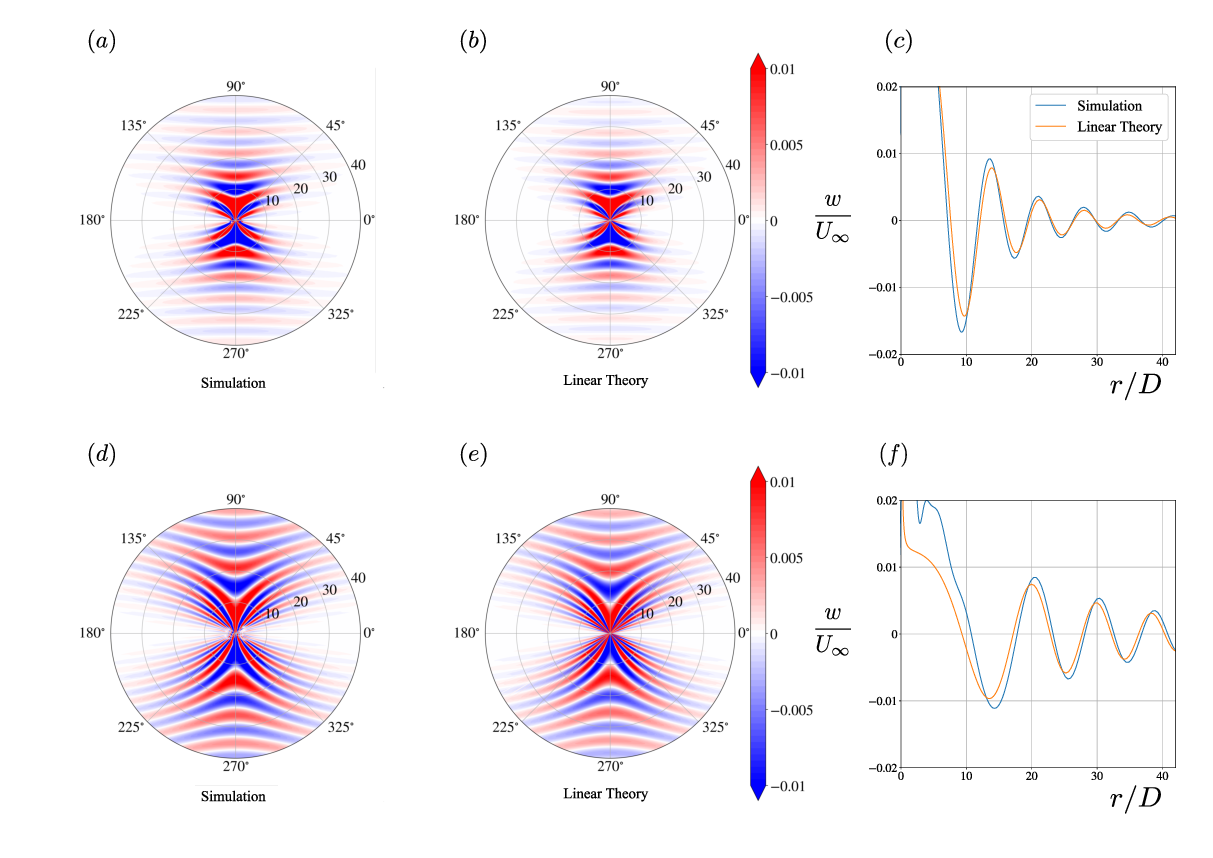}}
\caption{{Body-generated lee waves shown at two downstream locations for $\Fro =1$: vertical velocity contours compared with linear theory at $x/D = 10 \, (a,b,c)$ and $30\, (d,e,f)$. Line plots compare $w$ at $\theta = 90^{\circ}$.}}
\label{fig:lw_strm_comp}
\end{figure}

\begin{figure}

\centerline{\includegraphics[width=0.9\linewidth, keepaspectratio]{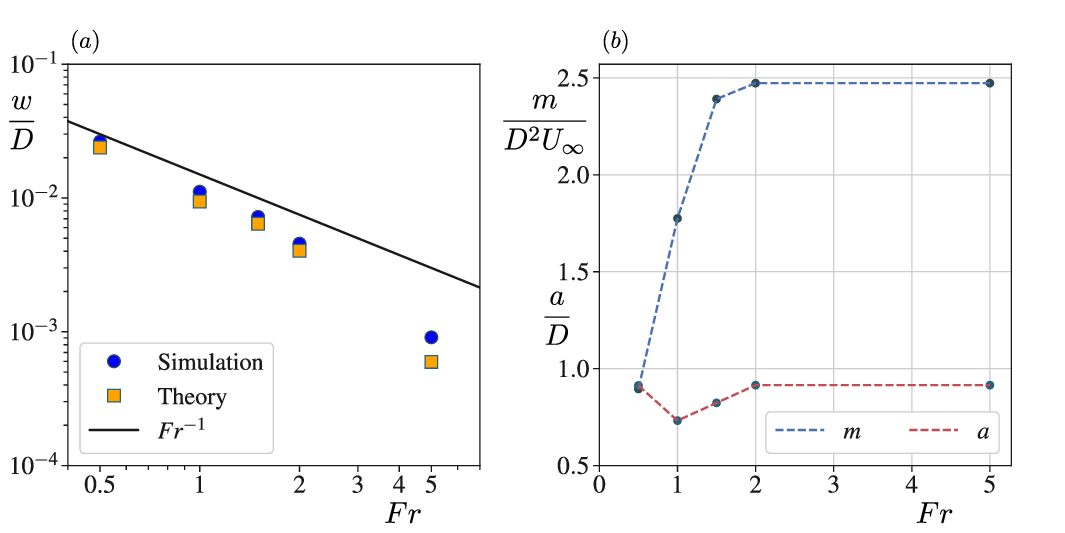}}
\caption{(a) Vertical mean velocity magnitude of the lee waves at local maxima/minima near $x/D = 20$ in part (a) of figure \ref{fig:separation_bubble};  {(b) Variation of $m$ and $a$ (obtained by fitting the resulting ovoid to the mean separation bubble)  with $\Fro$.}}
\label{fig:m_vs_fr}
\end{figure}

\begin{figure}

\centerline{\includegraphics[width=0.9\linewidth, keepaspectratio]{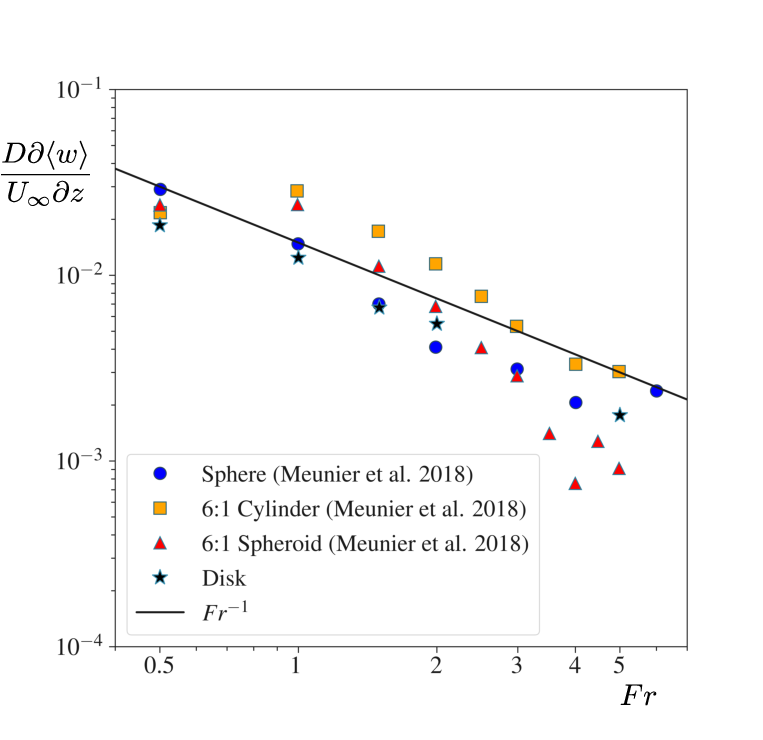}}
\caption{ Comparison of mean divergence magnitude of the lee-wave field between the present disk simulation and the laboratory results reported by \cite{meunier_internal_2018}.}
\label{fig:dwdz_comparison}
\end{figure}

For the steady lee waves, \cite{gola_pycnocline_2023}  found that the wavelength in all cases and the amplitude in the linearly-stratified  $\Fro =1$ case could  be deduced from the linear theory (with the value of $N$ in the far field where waves propagate) given by \cite{voisin_internal_1_1991, voisin_iwb_2003, voisin_sphere_2007} and by modeling the disk plus its separation bubble as one body, specifically a Rankine ovoid. The potential flow past an ovoid is given by the superposition of a doublet (source/sink of strength $\pm m$ separated by distance of $2a$) and the uniform incident flow. The expression for the vertical velocity deduced by \cite{ortiz_spheroid_2019} after adapting the linear theory to an ovoid-shaped body is 

\begin{equation}
w(x,y,z) {\sim}  -\frac{mN}{\pi U_{\infty} r_{xyz}} \sin{\theta} \cos{\psi} (1+\cot^{2}{\psi}\cos^{2}{\theta})^{\frac{1}{2}} \sin{\Big(\frac{Na}{U_{\infty}} \cos{\psi} |\sin{\theta}| \Big)} \sin{\Big(\frac{N}{U_{\infty}} r_{xyz} |\sin{\theta}| \Big)} ,
\label{w_lee_waves}
\end{equation}
 where $r_{xyz}=\sqrt{x^{2} + y^{2} + z^{2}}$, $\psi = \arctan(\sqrt{y^{2}+z^{2}}/x)$ and $\theta = \arctan{(z/y)}$ {is the azimuthal angle in figure \ref{fig:simsetup}}. 
 Specialization to the $y = 0$ plane leads to
 \begin{equation}
w(x,y=0,z) {\sim}  -\frac{mN}{\pi U_{\infty} r_{xz}} \cos{\psi} \sin{\Big(\frac{Na}{U_{\infty}} \cos{\psi}\Big)} \sin{\Big(\frac{N}{U_{\infty}} r_{xz}\Big)} ,
\label{w_lee_waves_1}
\end{equation}
 where $r_{xz}=\sqrt{x^{2}+z^{2}}$ and $\psi = \arctan(z/x)$. The theoretical result, (\ref{w_lee_waves_1}), will be  subsequently compared with the simulation results.
 
 As in the $\Fro =1$ case tested by \cite{gola_pycnocline_2023}, $m$ and $a$ are calculated here  from the potential-flow solution for a Rankine ovoid of length $L_{ro}$ and cross-sectional diameter $D_{ro}$ {fitted using the streamline $\langle u \rangle = 0.95 U_{\infty}$}:
\begin{equation}
(L_{ro}^2 - 4a^2 )^2 = \Big(  \frac{8ma}{\pi U_{\infty}}  \Big) L_{ro} \qquad, \qquad D_{ro}^2 = \Big(  \frac{8ma}{\pi U_{\infty}}  \Big) \frac{1}{\sqrt{4a^2 + D_{ro}^2}}
\label{lro_dro}
\end{equation}

It can be deduced from the argument of the last sine term in {(\ref{w_lee_waves_1})} that the wavelength of the steady lee waves is $\Lambda / D = 2\pi \Fro$. The wavelength of the lee waves obtained from the simulations along the line $x=z, y=0$ is shown in figure \ref{fig:lw_wl}, wherein it follows the trend $\Lambda = 6.7 \Fro$, which is close to the asymptotic result of $\Lambda / D = 2 \pi \Fro$.

{Note that (\ref{w_lee_waves}) does not apply to $\Fro < O(0.1)$ when the flow is mostly around the disk and the region involved in wave generation shrinks from the crest to the dividing streamline at a  depth of approximately $U_\infty/N$ from the crest. Thus, the singular limit $\Fro \rightarrow 0$ of (\ref{w_lee_waves})  is inadmissible.}

Figure \ref{fig:separation_bubble} shows the vertical velocity $w$ for the five cases  on the line $z = x, y =0$ as a function of distance from the disk center, $s = \sqrt{x^{2} + z^{2}}$.  Excellent agreement of the simulation results with  linear theory, (\ref{w_lee_waves_1})), is seen.  Also plotted alongside in  figure \ref{fig:separation_bubble} (b)  are the {defect velocity ($U_{d} = U_{\infty} - \langle u \rangle$)} contours. There is a case-dependent difference in the separation bubble.  The length of  the separation bubble increases with increasing $\Fro$ leading to a different set of values for $m$ and $a$ in (\ref{w_lee_waves}) for each case. {Figure \ref{fig:lw_strm_comp} shows the comparison of the vertical velocity contours on the $r-\theta$ plane obtained from the simulation and linear theory (\ref{w_lee_waves}) for $Fr = 1$ at $x/D = 10$ and $30$. The contours are seen to span a larger area  in the $r-\theta$ plane as distance from the body is increased. Again, good agreement between the simulation and theory is found.} {The amplitude of the lee wave {(maximum or minimum at $x/D = 20, r/D > 10$), which is plotted in figure \ref{fig:m_vs_fr} (a), reveals excellent agreement between theory and simulation.} {The  $w$-amplitude of the lee wave in (\ref{w_lee_waves}) is proportional to the product of $m$ and $N$ so that the change in the size of the separation region behind the disk also plays a part in the resulting wave amplitude as is reflected by the dependence of $m$ in (\ref{w_lee_waves}) on $\Fro$ (Figure \ref{fig:m_vs_fr}b). This results in a scaling that is different from $w/U_{\infty} \sim{\Fro}^{-1}$, which would be the case if $m$ were assumed constant across \Fro.}

\cite{meunier_internal_2018} obtained the mean horizontal divergence magnitude ($|\partial \langle w \rangle / \partial z|$)  of the internal wave field in an experimental study of waves emitted by towing a sphere, 6:1 cylinder and 6:1 spheroid in a stratified fluid. Measurements of $u(x,y)$ and $v(x,y)$ on a horizontal plane were used to infer $|\partial \langle w \rangle / \partial z|$. For comparison, the mean horizontal divergence magnitude at the local maxima/minima for $w$ at $x/D = 10, r/D = 3$ for the disk simulations is plotted along with the data from \cite{meunier_internal_2018} in Figure \ref{fig:dwdz_comparison}.  Note that the  chosen locations of the maxima/minima in the simulations are somewhat different from those in \cite{meunier_internal_2018}, reflecting the difference between simulation and experimental methodologies. {The simulation results for the disk are comparable to the experimental results of \cite{meunier_internal_2018} for other axisymmetric bodies.}

\section{Wake generated internal gravity waves}\label{unsteady_waves}

\begin{figure}

\centerline{\includegraphics[width=0.9\linewidth, keepaspectratio]{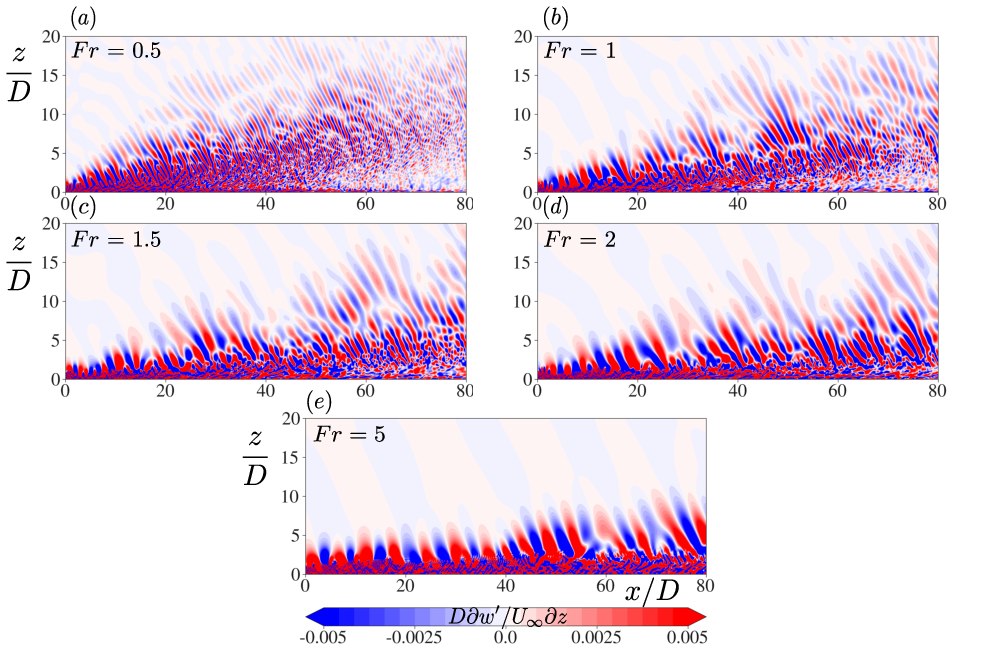}}
\caption{Instantaneous contours of vertical derivative of vertical velocity fluctuation showing wake generated internal gravity waves on the $\theta = 90^{\circ}$ plane for the five cases. 
}
\label{fig:unsteady_waves}

\end{figure}

The wake generated internal waves are manifested in the simulation reference frame as the unsteady component of the wave field.  To visualise these waves without their steady counterpart, instantaneous contours of the fluctuation ({$\partial w' / \partial z$}) in vertical divergence on the top-half vertical plane is plotted in Fig. \ref{fig:unsteady_waves} for all the five cases. In a time series of these contours, these waves are seen to advect downstream  {(see supporting video `Movie1.mp4')}, unlike the steady lee waves. When viewed in a reference frame moving with the freestream {(see supporting video `Movie2.mp4')}, these waves move  slowly towards the left while growing in the vertical. {In other words, the wave phase speed is less than the freestream velocity magnitude.}

\begin{figure}

\centerline{\includegraphics[width=0.9\linewidth, keepaspectratio]{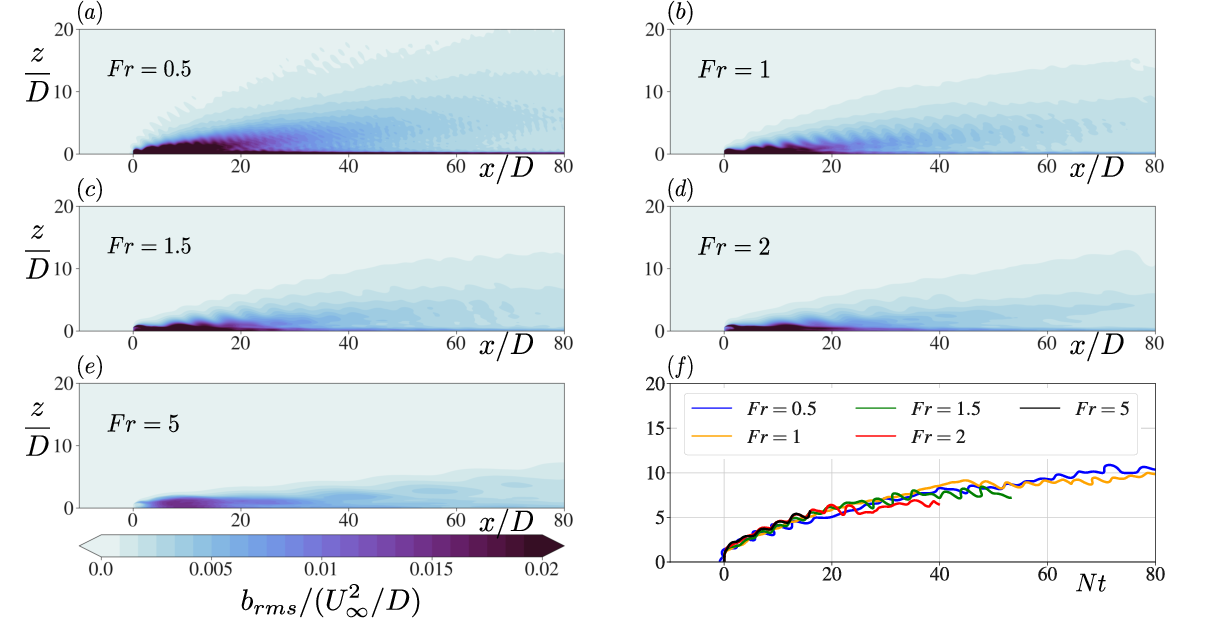}}
\caption{(a-e) Wake wave envelope visualised by the rms of buoyancy term for the five cases. (f) Contour lines of $b_{rms} = 0.0015 U_{\infty}^{2}/D$ plotted with $Nt$}
\label{fig:igw_env}

\end{figure}

Across the five cases, three features of these internal waves can be noted: (1) The inclination angle with the vertical ($\Theta$) is in a narrow range ($35^{\circ} - 40^{\circ}$), specifically $35^{\circ}, 39^{\circ}, 35^{\circ}, 40^{\circ}$ and $37^{\circ}$ in the order of increasing \Fro, (2) the wavelength increases with increasing \Fro,  (3) the vertical extent of propagation with respect to the body decreases with increasing \Fro. The first feature, the clustering of the waves around the same vertical angle, was explained using the argument of maximising the vertical component of group velocity for a fixed horizontal wavenumber by \cite{taylor_internal_2007}. The vertical group velocity for internal gravity waves in stratified flow without rotation is given by

\begin{figure}

\centerline{\includegraphics[width=0.85\linewidth, keepaspectratio]{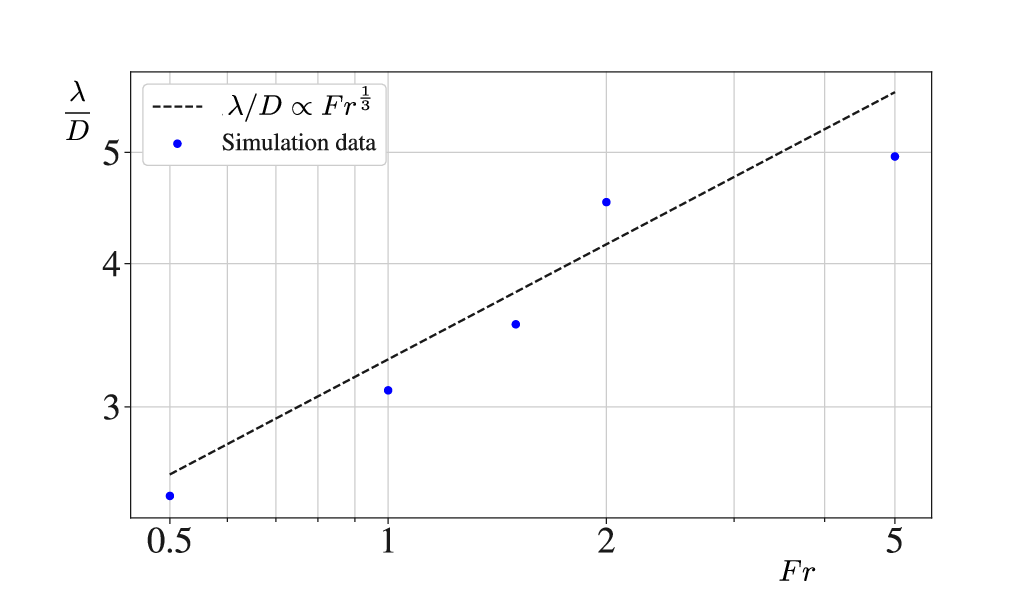}}
\caption{Wavelength of wake generated internal gravity waves. }
\label{fig:igw_wl}

\end{figure}

\begin{equation}
c_{gz} = -\frac{N}{|\vec{k|}} \cos{\Theta} \sin{\Theta} = -\frac{N}{|k_{x}|} |\cos{\Theta}| \cos{\Theta} \sin{\Theta},
\label{cg_z}
\end{equation}
where $|k|$ is the wavenumber magnitude and $k_{x}$ is the horizontal wavenumber. Note that for our simulations, $\Theta$ is an obtuse angle because the waves propagate in the negative $x$ direction as discussed earlier. It can be shown that the expression on the right hand side of  (\ref{cg_z})  is maximum when $\Theta \approx 145^{\circ}$ or $35^{\circ}$ from the vertical, which is close to the angles observed in the simulations.

It is evident from figure \ref{fig:unsteady_waves} that the wavelength of the wake waves increases with \Fro. This dependence is quantified by computing the wavelength ($\lambda$), normalized by $D$, from wavepackets  in the instantaneous $\partial w' / \partial z$ fields shown in figure \ref{fig:unsteady_waves}. The wavepackets are chosen to be $4-5$ wavelengths long in the NEQ region away from the wake to avoid turbulence but also close enough to get a strong clean signal (this corresponds to the region $5 < z < 15, 5 < Nt < 30$ depending on \Fro). {Figure \ref{fig:igw_wl} shows $\lambda$ plotted against $\Fro$. There is reasonable agreement with  the scaling $\lambda \propto \Fro^{1/3}$ (dashed line) found by \cite{abdilghanie_internal_2013, meunier_internal_2018}. The horizontal width ($L_{Hk}$) of the  kinetic energy profile of the disk wake varies as $L_{Hk} \propto x^{1/3}$ \citep{chongsiripinyo_decay_2020}. Assume that the energetic eddies  responsible for wave generation scale with $L_{Hk}$  and that the internal wave radiation becomes significant only when $Nt$ reaches a specific value. Then, the distance at which the internal wave radiation commences scales as $\Fro$, the value of  $L_{Hk}$  at that location varies as $\Fro^{1/3}$ and, therefore,  $\lambda \propto \Fro^{1/3}$. Indeed,  it will be shown in \S \ref{sec:energy_ratios} that the wake-wave potential energy (as a fraction of turbulent kinetic energy of the wake) becomes significant at  $Nt \approx 5$ and then does not change appreciably until  $Nt \approx 30$ when it decreases. 
}

Lastly, to explain the trend in vertical extent of the waves, the contours of the rms buoyancy term $b_{rms} = g \rho_{d_{rms}} / \rho_{0}$ are plotted in figure \ref{fig:igw_env} (a-e), thereby providing a good visualisation of the envelope of the wake generated internal waves. When the vertical extent of this envelope (quantified by a small value of $b_{rms} = 0.0015 U_{\infty} ^{2} / D$) is plotted against $Nt$ in figure \ref{fig:igw_env} (f), a good collapse is seen, showing  that the vertical extent of propagation is the same for these waves in buoyancy time units. For example, $x = 40$ corresponds to $Nt = 80$ for $Fr = 0.5$ and $Nt = 8$ for $Fr = 5$, accounting for the difference in wave envelope between these cases at the same $x$.

\section{Wake and wave energetics}\label{energetics}

A turbulent stratified flow has both kinetic and potential energy with each having a mean and a fluctuating component. Furthermore, the mean and the fluctuation can be divided into a turbulent wake component and a wave  part surrounding the wake component. In this section, the variation of the energy partition among its  components will be diagnosed. It will be shown that,  across the five levels of stratification,  there is a systematic variation of mean relative to turbulent components as well as potential versus kinetic energy.  The definition of various energy components is as follows:

Mean kinetic energy (MKE):

\begin{equation}
MKE = \frac{1}{2} ((\mean{u}-U_{\infty})^{2} + \mean{v}^{2} + \mean{w}^{2})
\end{equation}

Mean potential energy (MPE):

\begin{equation}
MPE = \frac{g^{2} \mean{\rho_d}^{2}}{2 \rho_{0}^{2} N^{2}}
\end{equation}

Turbulent kinetic energy (TKE):

\begin{equation}
TKE = \frac{1}{2} (\mean{u'^{2}} + \mean{v'^{2}} + \mean{w'^{2}})
\label{tke_def}
\end{equation}

Turbulent potential energy (TPE):

\begin{equation}
TPE = \frac{g^{2} \mean{\rho_d'^{2}}}{2 \rho_{0}^{2} N^{2}}
\label{tpe_def}
\end{equation}

The various plots in this section compare the evolution of these energy components,  integrated over the $r - \theta$ plane,  as a function of streamwise distance ($x$). Let $A$ be the complete $r-\theta$ plane in the simulation domain at any $x$. If $C$ is a closed curve on $A$ enclosing the area $A_{C}$, the following three integrals of a quantity $E(r, \theta, x)$ can be defined as a function of $x$ ({see figure \ref{fig:ellipse} for two examples of the curve $C$ chosen to separate the wave from the wake field}):

Total integral over the complete $r-\theta $ plane:

\begin{equation}
E_{TI}(x) = \int_{A} E(r, \theta, x) r \,dr \,d\theta,
\label{ti}
\end{equation}

inside integral over the area bounded by $C$ on $r-\theta $ plane (note that $C$  varies as a function of $x$):

\begin{equation}
E_{II}(x) = \int_{A_{C}(x)} E(r, \theta, x) r \,dr \,d\theta,
\label{ii}
\end{equation}

and outside integral, which is computed over the area external to that bounded by $C$:

\begin{equation}
E_{OI}(x) = \int_{A - A_{C}(x)} E(r, \theta, x) r \,dr \,d\theta,
\label{oi}
\end{equation}

so that, $E_{TI}(x) = E_{II}(x) + E_{OI}(x)$.
\subsection{Mean kinetic energy (MKE) and mean potential energy (MPE)}

\begin{figure}

\centerline{\includegraphics[width=0.9\linewidth, keepaspectratio]{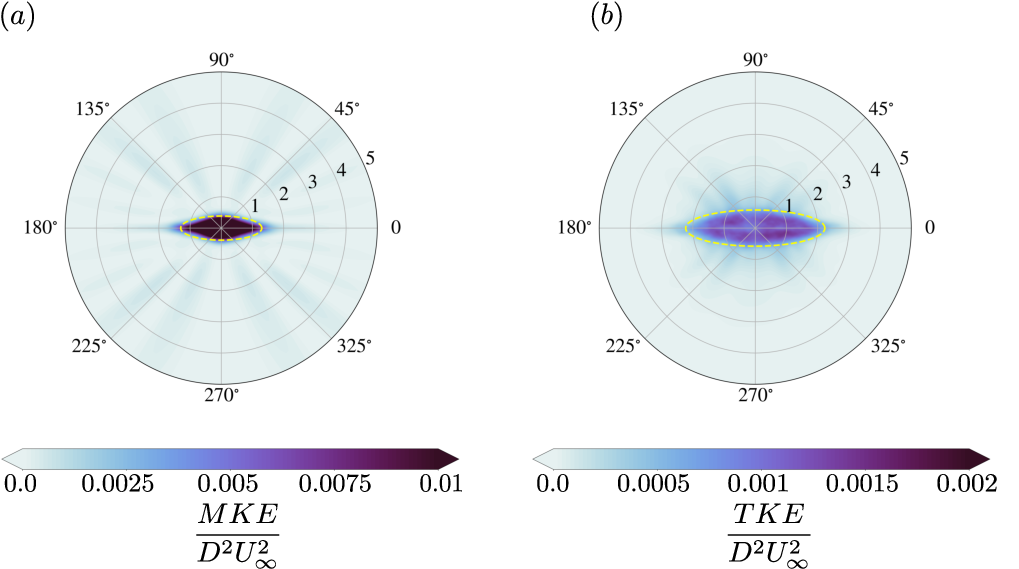}}
\caption{{Examples of curve $C$ (here an ellipse shown in dashed yellow line) used to separate the wave from the wake contribution: (a) MKE contours for $Fr = 1.5$ at $x/D = 20$ as separated by $C$ based on half width of wake $U_{d}$ profile, (b) TKE contours for $Fr = 1.5$ at $x/D = 20$ as separated by $C$ based on half width of wake TKE profile.} 
}
\label{fig:ellipse}

\end{figure}

\begin{figure}

\centerline{\includegraphics[width=0.9\linewidth, keepaspectratio]{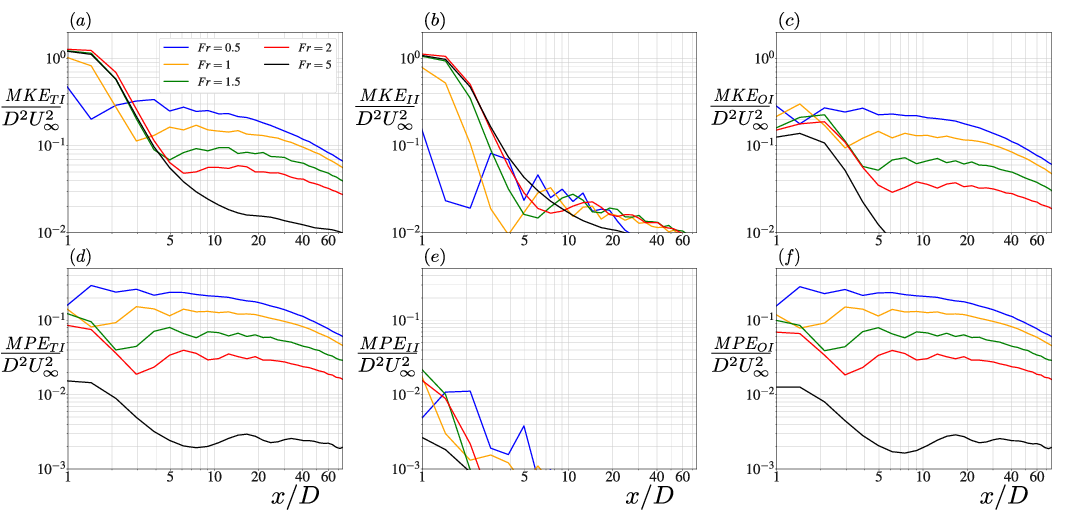}}
\caption{Area-integrated mean energies for all five cases. Total integral (TI) of the fluctuating energy is shown as well as its partition into wake or inside integral (II) and wave or outside integral (OI).
}
\label{fig:mpe_mke}

\end{figure}

The curve $C$ in (\ref{ii}), and (\ref{oi}) will be chosen to be at the edge of the wake so that the outside integral (OI) serves as a surrogate for the wave energy and the inside integral (II) for the wake energy.
Specifically,  an ellipse with its semi-major and semi-minor lengths equal to the horizontal and vertical half width of the mean defect velocity profile is chosen for each $\Fro$ case; these lengths  vary as a function of $x$. Thus, using equations (\ref{ti}), (\ref{ii}) and (\ref{oi}), the wake and wave contributions based on the inside and outside regions of the ellipse can be calculated.

The top row of figure \ref{fig:mpe_mke} shows the various integrals of $MKE$ for all five cases. For cases with lower $\Fro$, $MKE_{TI}$ (figure \ref{fig:mpe_mke}a) is initially low because of  shorter separation zones (see figure \ref{fig:separation_bubble}). 
However, $MKE_{TI}$ also exhibits slower decay  with increasing $x$ for the higher-$\Fro$ cases, reflecting the fact that stratification prolongs wake lifetime, so that the  $\Fro$ variation in $MKE_{TI}$ eventually reverses at later time. There is significant MKE  in the outside wave region (\ref{fig:mpe_mke}c), which can be attributed to the lee waves, except at the highest simulated $\Fro = 5$ case with  weak lee waves. Moving to MPE shown in the bottom row, comparison of  \ref{fig:mpe_mke}(e) and (f) shows that  the outside contribution ($MPE_{OI}$ ) owing to  steady lee waves  dominates over the wake contribution  {($MPE_{II}$)} over the entire streamwise extent of the flow. Evidently,  the turbulent wake does not lead to significant {\em mean} distortion of the isopycnals although there is fluctuating distortion leading to TPE as will be shown shortly. Beyond $x = 10$, both $MKE_{TI}$ and $MPE_{TI}$ decrease with increasing $\Fro$. The decrease is quite  substantial from $\Fro = 2$ to $\Fro = 5$. Note that the energy values are normalized using $U_\infty$ and $D$. Thus if the increase of $\Fro$ is due to a higher flow speed under the same background $N$, the $Fr$-related decrease in the dimensional value of energy would be less.
 
{To summarize, the distribution of mean energy associated with stratified flow past a disk  is such that almost all of the mean potential energy component resides outside the turbulent wake in the lee wave field. At \Fro = 2 or less, most of mean kinetic energy resides in the lee wave field in contrast to the higher \Fro = 5  case where it is the wake that carries most of the mean kinetic energy.}

\subsection{Turbulent kinetic energy (TKE) and turbulent potential energy (TPE)}

\begin{figure}

\centerline{\includegraphics[width=0.9\linewidth, keepaspectratio]{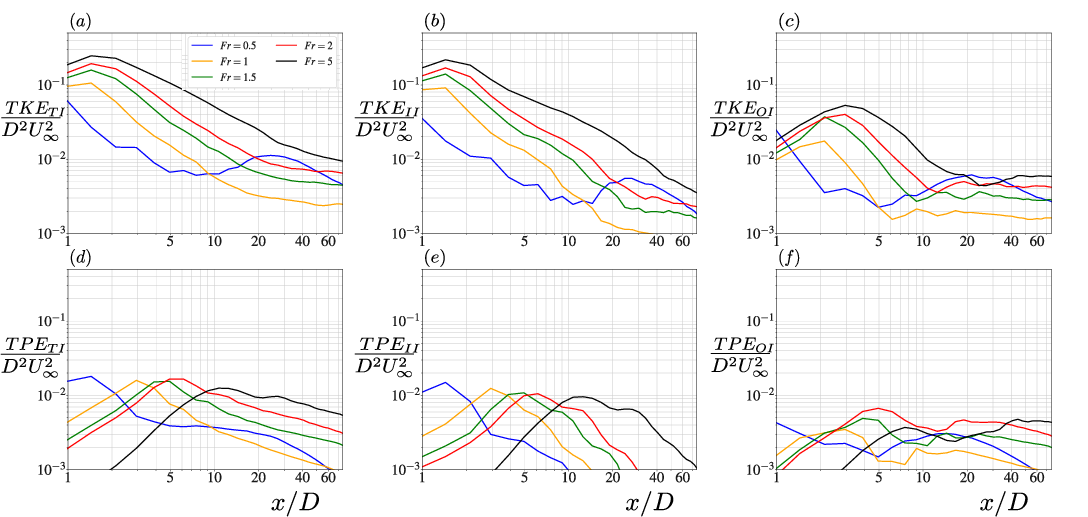}}
\caption{Area integrated fluctuation energies for all five cases. Total integral (TI) of the fluctuating energy is shown as well as its partition into wake or inside integral (II) and wave or outside integral (OI).}    

\label{fig:tpe_tke}

\end{figure}

Figure \ref{fig:tpe_tke} shows the various fluctuation energies for all five cases. The fluctuation energy can be divided into the turbulent energy of the wake and the outside fluctuation energy carried by wake generated internal waves. This is achieved by defining an ellipse with its semi-major and semi-minor axes equal to the horizontal and vertical halfwidth of the TKE profile, respectively. The various contributions to the turbulent component are plotted similar to what was shown for the mean component in figure \ref{fig:mpe_mke}. Looking at the $TKE_{TI}$ profiles in figure \ref{fig:tpe_tke}(a), it is evident that the general \Fro-dependence seen in figure \ref{fig:mpe_mke}(a) has been flipped, i.e. the total $TKE$ is larger for the  cases with larger \Fro. Furthermore, the cases with higher $\Fro$  also have stronger turbulence as shown by the interior integral $TKE_{II}$ in  figure \ref{fig:mpe_mke}(b). The exception is  the case of $Fr = 0.5$. Here, both $TKE_{II}$ (interior wake turbulence) and  $TKE_{OI}$ (external to the wake)  flatten at $x \approx 20, Nt \approx 40$ and then increase slightly before decaying again. The increase of  wake fluctuation energy in the late NEQ/early Q2D regime of the \Fro = 0.5 case is a result of the wake structure becoming increasingly two-dimensional and its flapping in the horizontal, similar to what was found for $\Fro < 1$ sphere wakes at $\Rey = 3700$  by \cite{Pal_JFMrapids2016}.

$TPE_{TI}$ for $Fr = 0.5$ also shows slight stalling at $x/D \approx 10$ before decaying again at $x/D \approx 30$, although this is likely the result of internal wave activity in the NEQ regime as can be seen from increase in external fluctuation energy, $TPE_{OI}$ in figure \ref{fig:tpe_tke}(f). 
Comparing figure \ref{fig:tpe_tke} with figure \ref{fig:mpe_mke} shows that $Fr = 5$ is qualitatively different than the other four cases, e.g., it has a larger potential energy in the fluctuations than in the mean.

\subsection{Ratio of kinetic and potential energies}
\label{sec:energy_ratios}

\begin{figure}

\centerline{\includegraphics[width=0.9\linewidth, keepaspectratio]{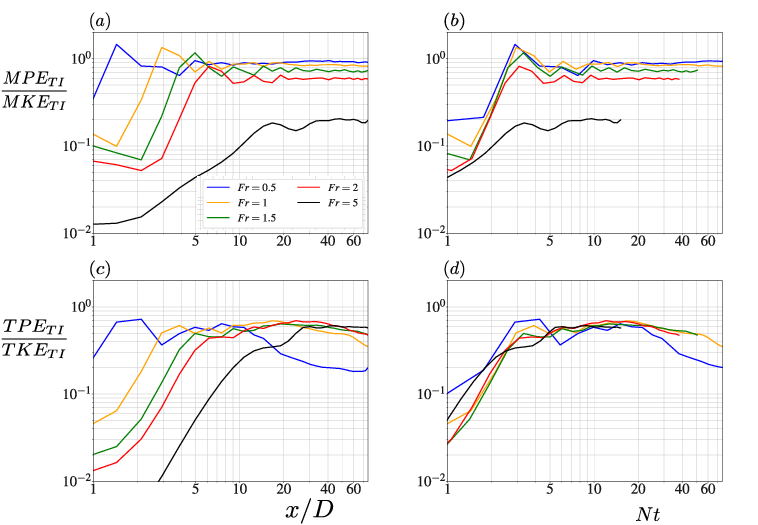}}
\caption{Ratio of  potential and kinetic energy  for all the five cases. This ratio is plotted for the mean and turbulent components as a function of $x/D$ (a and c) and $Nt$ (b and d).} 
\label{fig:energy_ratio}

\end{figure}

{Figure \ref{fig:energy_ratio} shows the mean and fluctuation energy ratios ($MPE_{TI}/MKE_{TI}$ and $TPE_{TI}/TKE_{TI}$), respectively, plotted  as a function of  $x/D$ as well as $Nt$. For the mean energies, the ratio $MPE_{TI}/MKE_{TI}$ approaches a constant value after $Nt \approx 5$. This constant $O(1)$ value has a small variation for the lower $\Fro$ cases, e.g between $0.9$ at $\Fro = 0.5$ to 0.6 at $\Fro = 2$.  However, there is a large decrease of the ratio to 0.2 at \Fro = 5, pointing to a decrease in the importance of the lee wave field at high \Fro. Insofar as  the fluctuation energy ratio,  $TPE_{TI}/TKE_{TI}$, it increases as the wake evolves to approach a plateau at  $0.6 \sim 0.7$ in the region $5 < Nt < 30$ (figure \ref{fig:energy_ratio}d). This plateau lies within the NEQ regime where the wake adjusts to the background stratification and the activity of the wake generated internal waves is the highest.
}

\subsection{Comparison of body  lee wave energy with wake wave energy }

\begin{figure}

\centerline{\includegraphics[width=0.9\linewidth, keepaspectratio]{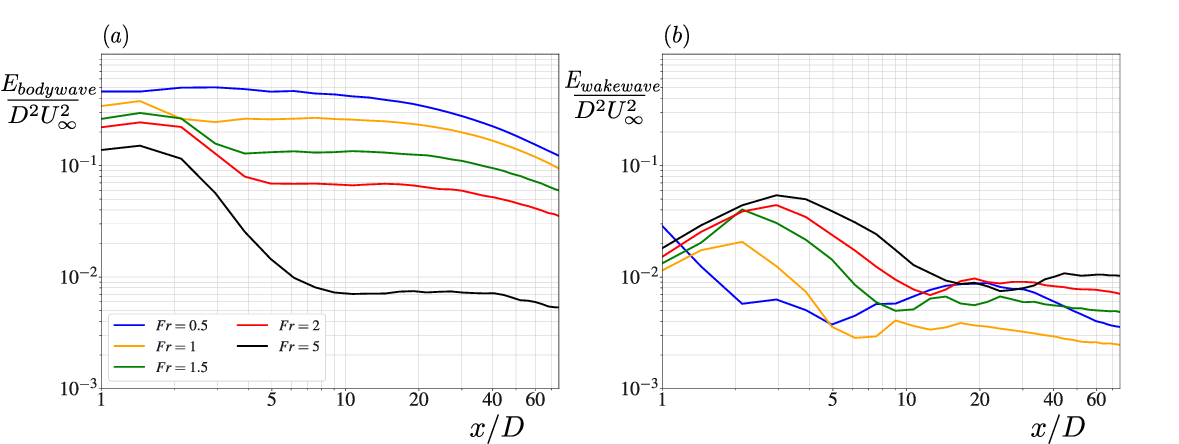}}
\caption{(a) Total energy in body lee waves (b) Total energy in wake generated internal waves.} 
\label{fig:bw_ww}

\end{figure}

Previous laboratory measurements~\citep{brandt_internal_2015, meunier_internal_2018} at selected spatial locations  have shown that, while the body wave field is more energetic than the  wake wavefield  at lower \Fro, there is a crossover to dominance of wake wave energy at sufficiently large \Fro.  The  wave energy is computed  from the outside integral terms, i.e. $E_{\rm body \,  wave} = MPE_{OI} + MKE_{OI}$  and $E_{\rm wake \,  wave} = TPE_{OI} + TKE_{OI}$.  Figure \ref{fig:bw_ww} compares the downstream evolution of $E_{\rm body \, wave}$  and $E_{\rm wake \,  wave}$. 
Unlike the behavior in the lower-$\Fro$ cases, the energy in the wake generated internal waves at at $\Fro = 5$ surpasses the energy in steady lee waves. Thus, in the disk wake, a crossover to the dominance of wake waves occurs at a value of $\Fro$ between 2 and 5.

\section{Comparison with disk wakes at $Re = 50,000$} \label{comparison}

In this section, internal waves  from  the data  \citep{chongsiripinyo_decay_2020} for a disk wake at a higher $\Rey = 50,000$ and with $\Fro = 2, 10$ are quantified and the results compared with the $\Fro$ dependencies discussed in the previous sections.

\subsection{Structure of body generated lee waves and wake generated internal gravity waves}

\begin{figure}

\centerline{\includegraphics[width=0.9\linewidth, keepaspectratio]{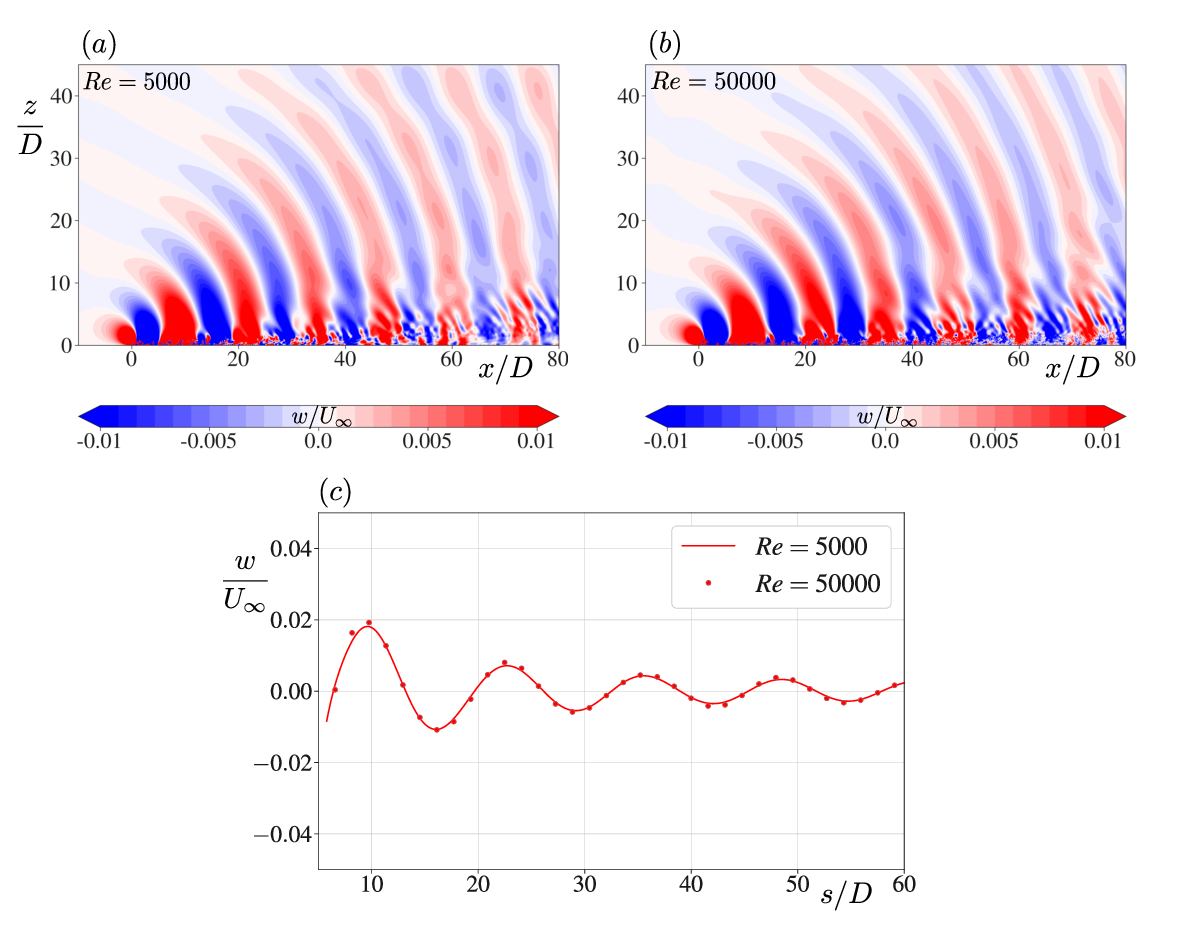}}
\caption{Lee wave comparison for $Re = 5000$ and $Re = 50000$ at $Fr = 2$: (a,b) Instantaneous vertical velocity contours, (c) Vertical velocity on the line $z=x, y=0$ ($s = \sqrt{x^{2} + z^{2}}$).}
\label{fig:lw_5k_50k}

\end{figure}

Lee waves for $\Rey = 5000$ and $\Rey = 50,000$ at $\Fro = 2$ are compared in figure \ref{fig:lw_5k_50k} by plotting vertical velocity ($w$) contours on the $\theta = 90^{\circ}$ plane and also the value of $w$ along the line $z=x,y=0$. The influence of $\Rey$ on the wavelength and spatial distribution of the wave field is negligible. The lee waves for the $\Rey = 50000$ case at $\Fro = 10$ (not shown here) are weaker than at $Fr =2$ and have longer wavelength, in agreement with the linear theory result of section \ref{lee_waves}.

\begin{figure}

\centerline{\includegraphics[width=0.9\linewidth, keepaspectratio]{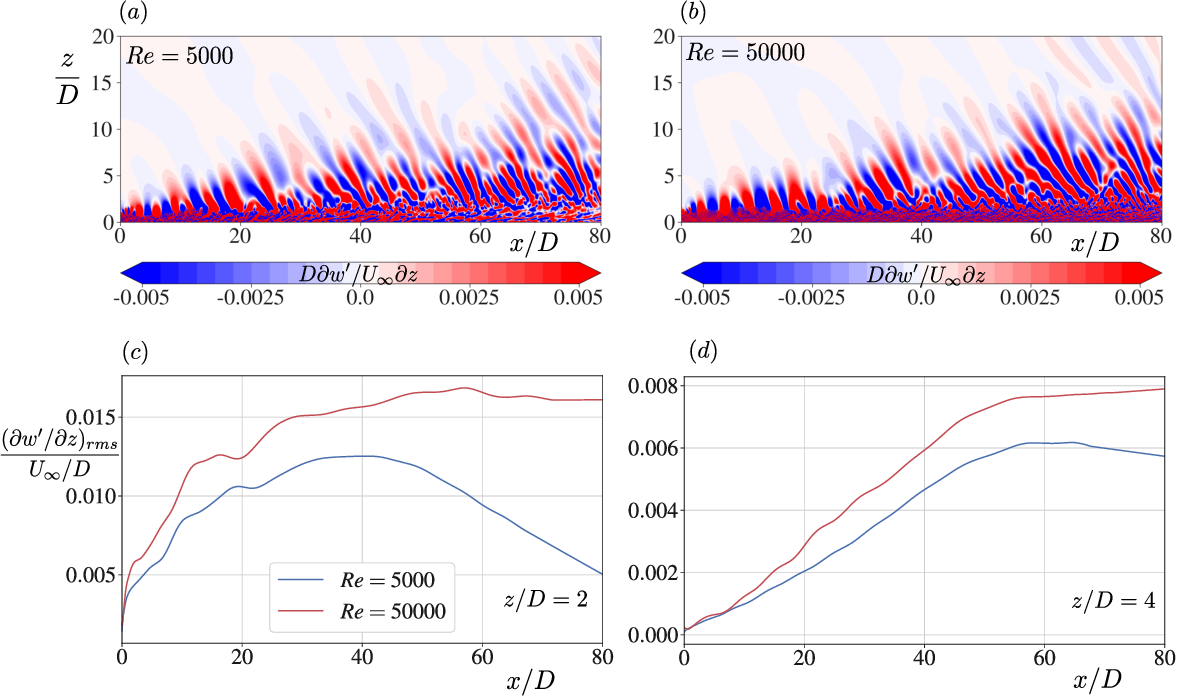}}
\caption{Wake generated internal wave comparison for (a) $Re = 5000$ and (b) $Re = 50000$ at $Fr = 2$.}
\label{fig:igw_5k_50k}

\end{figure}

The influence of $\Rey$ on  wake waves is illustrated by figure \ref{fig:igw_5k_50k} (a,b) that shows $\partial w' / \partial z$ on the vertical center half plane ($\theta = 90^{\circ}$). Similarity is observed in the inclination of the waves ($\Theta \approx 40^{\circ}$) as well as their wavelength ($\lambda/D \approx 4.5$). The $\Rey = 50,000$ wake is more turbulent than at $\Rey = 5000$ as suggested visually by figure \ref{fig:igw_5k_50k} (b) with  enhanced small-scale features in the wake core relative to \ref{fig:igw_5k_50k} (a). 

The wake wave amplitude, measured by $(\partial w' / \partial z)_{rms}$, is compared between $\Rey = 5000$  and $\Rey = 50,000$ in figure \ref{fig:igw_5k_50k} (c,d). Proximity to the wake leads to a higher difference in wave amplitude, as is observed at $z/D = 2$ relative to $z/D = 4$. The scales of wake structures that give rise to wake waves are more broadband for $\Rey = 50,000$ as compared to $\Rey = 5000$ and at higher $\Rey$, waves are less affected initially by viscosity.

\cite{meunier_internal_2018} report an empirical scaling of $\Rey^{0.4}$ for the wake wave amplitude and \cite{abdilghanie_internal_2013} report that wave momentum flux varies as  $\langle uw \rangle \sim \Rey^{0.25}$. Since the increase of wave amplitude with $\Rey$ depends on streamwise location in the present simulations, we refrain from proposing a power law dependence on $\Rey$.

\subsection{Plane-integrated wave energy }

\begin{figure}

\centerline{\includegraphics[width=1.1\linewidth, keepaspectratio]{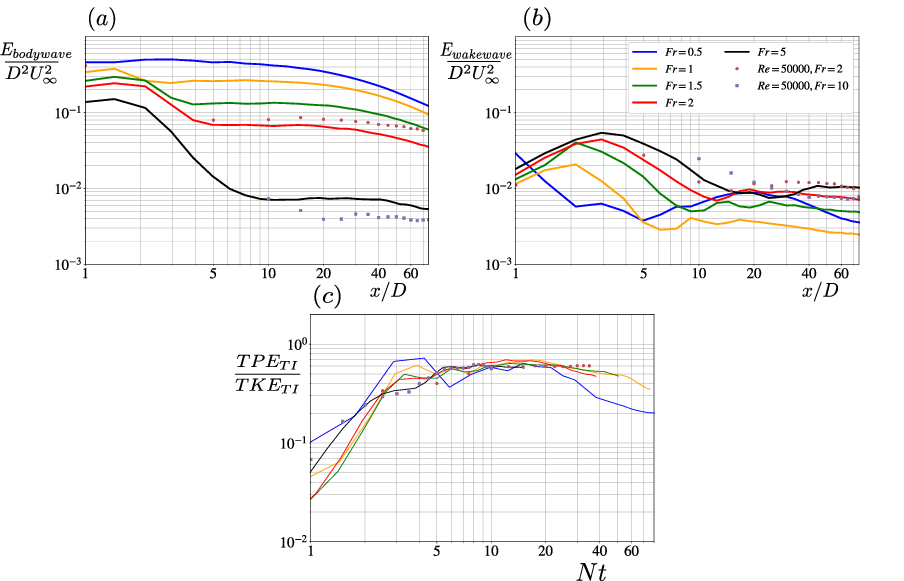}}
\caption{(a) Total energy in body lee waves (b) Total energy in wake waves (c) Ratio of total turbulent potential energy and total turbulent kinetic energy.}
\label{fig:ec_5k_50k}

\end{figure}

Analysis of the $\Rey = 5000$ results showed 
that when $Fr$ is increased, the total energy in wake waves progressively increases while the total energy in body waves decreases. Those results also showed that at  $Fr = 5$,  the wake wave energy  overtook  the body wave energy. Figure \ref{fig:ec_5k_50k}(a,b) shows the downstream variation of body wave energy ($E_{\rm body \, wave} = MPE_{OI} + MKE_{OI}$) and 
wake wave energy ($E_{\rm wake \, wave} = TPE_{OI} + TKE_{OI}$),  after including the additional results  for $\Rey = 50,000$ at $Fr = 2, 10$ to the $\Rey = 5000$ cases. The streamwise evolution of  $E_{\rm body \, wave}$ at $\Fro = 2$ is qualitatively  similar between the two values of $\Rey$  although the values are slightly larger at the higher $\Rey$.  The $\Fro = 10$ case in figure \ref{fig:ec_5k_50k}(a) has the weakest stratification and also the smallest body wave energy among all the simulations, which is consistent with the trend at $\Rey = 5000$. Turning to the wake wave energy as a function of $x/D$, the $\Fro = 2 $ case shows consistently higher wave energy at $\Rey = 50,000$ relative to the lower-$\Rey$ case.

When the ratio $TPE_{TI}/TKE_{TI}$ is plotted against $Nt$ in figure \ref{fig:ec_5k_50k} (c), now also including the data for $Re = 50000$, the two new cases also show that the ratio increases as a function of $Nt$ until reaching a maximum value of $0.6 - 0.7$ in the NEQ region,  $5 < Nt< 30$. This implies that the balance between the turbulent energies for a stratified flow evolves in a similar fashion not only across different $Fr$ for a fixed $Re$, but also for higher $Re$, pointing towards a potential universality of this statistic for  stratified flows.

\section{Summary and Discussion} \label{conclusions}

Body-inclusive large eddy simulations are used to study the effect of changing stratification levels in   flow past  a disk. The focus is on internal gravity waves. { Systematic investigation of internal waves over a range of $\Fro$  is new insofar as   body-inclusive simulations and adds to our knowledge base from previous experiments, temporal-model simulations and theory.} Linear stratification at five different  levels is considered: $\Fro = 0.5, 1, 1.5, 2$ and $5$ and the flow is quantified up to a relatively long streamwise distance of $x/D = 80$. The Reynolds number based on the freestream velocity ($U_\infty$) and disk diameter ($D$) is fixed at $5000$ and, thus, variability across these cases can be attributed to $\Fro$ alone. The dataset of \cite{chongsiripinyo_decay_2020} for disk at $\Rey = 50,000$ at $Fr =2,10$ is also analysed for studying Reynolds number dependence. 

The questions posed in section \ref{objectives} are answered and discussed below:

\subsection{Prediction of lee waves}

With regards to the steady lee waves, the nondimensional amplitude ($w$) of the vertical velocity decreases with increasing \Fro.  All  five cases showed excellent agreement with the linear theory for steady lee waves by \cite{voisin_sphere_2007} when an equivalent body including the separation bubble was used to calculate the potential flow solution. As to the separation bubble, its length  differed considerably among the $\Fro = 0.5, 1$ and $1.5$ cases but not so much between  $\Fro = 2$ and $\Fro = 5$. {According to linear theory (\ref{lro_dro}), $w/ U_\infty \propto Fr ^ {-1}$ if $m$ is constant and the change of $m$ adds some variability.} The wavelength ($\Lambda$) was found to increase linearly with $\Fro$ according to {$\Lambda / D \sim 6.7 \Fro$}, close to the asymptotic relation of $\Lambda / D = 2\pi \Fro$. 
{ The amplitude of mean $\partial w /\partial z$, which was obtained by \cite{meunier_internal_2018} in laboratory experiments, was compared to the present simulation results and good mutual agreement was found.  The theory proposed by \cite{meunier_internal_2018} involves a forcing of the momentum conservation equation (to model the drag) as the leading-order contribution to wave generation while the theory of  \cite{voisin_sphere_2007} that was adapted for the present application is different and involves a volumetric forcing of the continuity equation. Our adaptation of the theory involves the inclusion of the separation bubble, whose size is related to and generally increases with the drag force, as part of the wave forcing. Thus, although not equivalent, both the theory used here and that used by \cite{meunier_internal_2018} lead to higher wave amplitude for blunter body shapes.}

\subsection{Properties of wake  waves}

 Unlike the  body-attached lee waves, the wake waves  advect downstream with respect to the disk. 
In a reference frame moving with the free stream, the wake waves move upstream consistent with the notion that the wake has a velocity deficit with respect to the freestream.
A narrow band ($\Theta = 35 - 40^{\circ}$ with respect to the vertical) is found for  the wave propagation angles  for all the five $\Fro$ cases. 
Narrow-band emission from turbulent shear flows has been found previously, e.g. from a shear layer (laboratory experiments by \cite{sutherland_internal_1998} and DNS by \cite{pham_dynamics_2009}), a turbulent Ekman layer  \citep{taylor_internal_2007}, and temporal-model simulations of a wake \citep{abdilghanie_internal_2013}. The wave propagation angles observed here are consistent with the analytical result of  $35^\circ$ that follows from the vertical group velocity maximisation (equivalently, viscous decay minimization) criterion given by \cite{taylor_internal_2007}. It is worth noting that \cite{abdilghanie_internal_2013} found that the narrow band  in their  lower $\Rey$ cases expanded to a wider band of inclination angles ($26-50^{\circ}$) at  higher $\Rey$; a possible reason is the reduced effect of viscosity on the waves. 

  Other characteristics of the wake waves were studied as well.  The wavelength of these waves increases with an increase in $\Fro$, agreeing  well with the $\lambda \propto \Fro^{1/3}$ scaling found by \cite{abdilghanie_internal_2013}  and \cite{meunier_internal_2018}. Also, the wavelength remains approximately constant until the end of the test domain at $x = 80$.  The  horizontal  width of the TKE profile increases as $x^{1/3}$ and the wave potential energy as a fraction of turbulent kinetic energy becomes significant at $Nt = Nx/U_\infty \approx 5$ for all five cases. Thus, the streamwise distance at which the internal wave burst  becomes significant scales as \Fro, the  length scale of the energetic flow structures (same order as the TKE profile thickness)  at that  point scales as $\Fro ^{1/3}$, and so does the corresponding wavelength of the radiated internal wave. 
  
   The envelope of the wake waves is quantified  by choosing the rms value of the buoyancy at the envelope boundary to be a small fixed  value. The width of the thus defined envelope at a given $x$ expands with decreasing \Fro.
  Nevertheless, self-similar evolution is found when the wave envelope is  plotted as as a function of the buoyancy time scale ($Nt$).  The use of $Nt$ in place of $x/D$  is necessary because the evolution time scale of wake waves is a buoyancy effect that depends on stratification ($1/N$) and not on advection ($D/U_{\infty}$).

\subsection{Wave and wake energetics}

Because of the presence of both a turbulent wake  and  
internal gravity waves external to the wake, it makes sense to partition the energies in the flow into their respective interior (wake) and  exterior (wave) contributions. This is done by calculating the area integrals of the energy which are then segregated into their wake and wave parts. The interior wake part of the energy  is defined to be inside an ellipse whose dimensions are determined by the $x$-dependent values of the  vertical and horizontal half-widths of the wake and the exterior wave part is the remaining portion. The wake/wave decomposition is applied to the kinetic energy (both the mean component MKE and the turbulent component TKE) and the potential energy (mean component MPE and turbulent component TPE).

{Comparison of the body lee wave energy (external MPE + MKE) and wake wave energy (external TPE + TKE) shows that lee waves are a dominant feature for lower $\Fro \sim O(1)$ while unsteady wake waves become more significant as $\Fro$ is increased. For the disk wake, $\Fro = 2$ is below the crossover and $\Fro =5 $ is above the cross-over point at which both types of waves have similar energetic importance. The crossover from lee-wave dominance to wake-wave dominance at sufficiently large $\Fro$ has been seen at selected measurement stations in   previous experimental studies, e.g.  \cite{brandt_internal_2015,meunier_internal_2018}.}

 For the mean flow, it is found that although MKE gets its contribution from both the wake defect and the lee waves, MPE is primarily in the lee waves with $\Fro = 0.5$ having the highest value. Beyond the near wake, the wake contribution to MKE increases with a decrease in $\Fro$, i.e. strengthening stratification, consistent with stratification prolonging the lifetime of the wake. The ratio of total MPE to total MKE reaches a case-dependent constant value beyond $Nt \approx 5$ that decreases with increasing \Fro. The ratio takes  higher values of 0.6 -1 at $ \Fro $ between 0.5 and 2 -  the regime where lee waves dominate over wake waves.
 
  For the fluctuation energies, $\Fro = 5$ has the highest wake TKE.
 Notably, the ratio of TPE and TKE exhibits a universal trend based on all the cases considered in this study, where it is seen to have a constant value of 0.6 to  0.7 for $5 < Nt < 40$, which coincides with the NEQ regime, {wherein the wake generated internal wave activity is the highest as part of the adjustment of  the  wake turbulence to the background stratification.}

\subsection{Effect of increasing Reynolds number}

{Using the dataset of \cite{chongsiripinyo_decay_2020}, it was established that the increase in Reynolds number has little  effect on  lee wave properties such as wavelength and amplitude. However, the amplitude of wake waves, measured by $(\partial w' / \partial z)_{rms}$, increases with $\Rey$. The relative enhancement of wave energy  is higher at locations closer to the wake. The increase of wake wave amplitude with increasing $\Rey$ is consistent with  previous results \citep{abdilghanie_internal_2013, meunier_internal_2018}. Importantly, the universality of the ratio of turbulent potential energy and turbulent kinetic energy is verified at the higher $\Rey = 50,000$ for both $Fr =2$ and $10$. The ratio has the value $0.6 \sim 0.7$ for all $\Rey$ and $Fr$ cases in this study.}

\subsection{Limitations and outlook}
Internal waves are examined in this study with body-inclusive simulations  for $\Fro$ between 0.5 and 10. It is of interest to expand the parameter space  to $\Fro > 10 $, a regime with initially weak  buoyancy effects,  and test the present scaling laws on wake wave structure and energetics. At such high values of \Fro, a large streamwise domain with  $x/D = \Fro Nt$ is required to capture the regime ($5 < Nt < 40$) that has energetic internal waves. The robustness of the present results at higher $\Rey$ also deserves further study. {A pure body-inclusive approach  with the long streamwise domains required at  high $\Fro $ or the fine grid spacing required at high $\Rey$ would incur excessively high computational cost.} The hybrid approach, which  combines a body-inclusive simulation with a body-exclusive  simulation (either spatial or temporal), is a good option to extend the $\Fro$ and $\Rey$ range of simulations.

\backsection[Funding]{The authors gratefully acknowledge the support of Office of Naval Research Grant N00014-20-1-2253.}

\backsection[Declaration of interests] {The authors report no conflict of interest.}

\bibliographystyle{jfm}

\bibliography{disk_re5k_igw_final}

\end{document}